\documentclass[sigconf]{acmart}

\usepackage{latexsym}
\usepackage{graphicx}
\graphicspath{{./images/}}
\usepackage{booktabs} 
\usepackage{color}  
\usepackage{amsmath}  
\usepackage{subcaption}
\usepackage{caption}
\usepackage{balance}
\usepackage{verbatim}
\usepackage{cancel}
\usepackage{tikz}
\usepackage{colortbl} 
\usepackage{framed}

\newenvironment{Snugshade}[1][236,236,236]{
    \setlength{\itemsep}{0pt}
     \setlength{\parsep}{0pt}
     \setlength{\topsep}{0pt}
     \setlength{\partopsep}{0pt}
     \setlength{\leftmargin}{1.5em}
     \setlength{\labelwidth}{0em}
     \setlength{\labelsep}{0em} 
\setlength{\parskip}{0pt}
    \definecolor{shadecolor}{RGB}{#1}%
    \begin{snugshade}%
}{%
    \end{snugshade}%
}

\acmConference[CIKM'18]{ACM CIKM Conference}{October 2018}{Turin, Italy}

\begin{document}

\title{TEQUILA: Temporal Question Answering over Knowledge Bases}

\author{Zhen Jia}
\affiliation{
    \institution{Southwest Jiaotong University, China}
}
\email{zjia@swjtu.edu.cn}

\author{Abdalghani Abujabal}
\affiliation{%
  \institution{MPI for Informatics, Germany}
}
\email{abujabal@mpi-inf.mpg.de}

\author{Rishiraj Saha Roy}
\affiliation{
  \institution{MPI for Informatics, Germany}
}
\email{rishiraj@mpi-inf.mpg.de}

\author{Jannik Str\"{o}tgen}
\affiliation{%
  \institution{Bosch Center for AI, Germany}
}
\email{jannik.stroetgen@de.bosch.com}

\author{Gerhard Weikum}
\affiliation{
  \institution{MPI for Informatics, Germany}
}
\email{weikum@mpi-inf.mpg.de}

\renewcommand{\shortauthors}{Z. Jia et al.}

\newcommand\BibTeX{B{\sc ib}\TeX}

\newcommand{\struct}[1]{\texttt{\small #1}}
\newcommand{\utterance}[1]{\textit{``#1''}}
\newcommand{\phrase}[1]{\textit{`#1'}}
\newcommand{\old}[1]{}

\newcommand{\squishlist}{
 \begin{list}{$\bullet$}
  { \setlength{\itemsep}{0pt}
     \setlength{\parsep}{3pt}
     \setlength{\topsep}{3pt}
     \setlength{\partopsep}{0pt}
     \setlength{\leftmargin}{1.5em}
     \setlength{\labelwidth}{1em}
     \setlength{\labelsep}{0.5em} } }

\newcommand{\squishend}{
  \end{list}  }

\newcommand{\comm}[1]{}

\begin{abstract}
Question answering over knowledge bases (KB-QA) poses challenges in handling 
complex questions that need to be decomposed into sub-questions.
An important case,
addressed here, is that of temporal questions, where cues for
temporal relations need to be discovered and handled. We present
\emph{TEQUILA}, an enabler method for temporal QA that can run on top of
any KB-QA engine.
TEQUILA has four stages. It detects if a question has temporal intent. It 
decomposes and rewrites the question into non-temporal sub-questions and
temporal constraints. Answers to sub-questions are then retrieved from
the underlying
KB-QA engine. Finally, TEQUILA uses constraint reasoning on temporal intervals to
compute final answers to the full question. Comparisons against state-of-the-art
baselines show the viability of our method.
\end{abstract}

\maketitle

\section{Introduction}
\label{sec:introduction}

\textbf{Motivation and Problem.} Knowledge-based question answering (KB-QA) aims 
to answer
questions over large knowledge bases (e.g., DBpedia, Wikidata, YAGO, etc.) or
 other structured data.
%
KB-QA systems take as input questions such as:
\begin{Snugshade}
Q1: \utterance{Which teams did Neymar play for?}
\end{Snugshade}
%
\noindent and translate them into structured queries, in a formal language
 like SPARQL or SQL,
%
\noindent and execute the queries to
retrieve answers from the KB.
In doing so, KB-QA methods need to address the vocabulary mismatch between
phrases in the input question and
entities, types, and predicates in
the KB:
mapping \phrase{Neymar} to the uniquely identified entity, \phrase{teams} to 
the KB type \struct{footballClub} and \phrase{played for} to the KB predicate
\struct{memberOf}. 
State-of-the-art KB-QA (see surveys~\cite{diefenbach:17,moschitti:17})
can handle simple questions like the above example very well, but 
struggle with complex questions that involve multiple conditions
on different entities and need to join the results from corresponding
sub-questions. 
For example, the question:
\begin{Snugshade}
Q2: \utterance{After whom did Neymar's sister choose her last name?} 
\end{Snugshade}
\noindent would require a three-way join that connects Neymar, his sister
Rafaella Beckran, and David Beckham.

An important case of complex questions are temporal information needs.
Search often comes with explicit or implicit conditions about
time~\cite{metzler:09}.
Consider the two examples:
\begin{Snugshade}
Q3: \utterance{Which teams did Neymar play for before joining PSG?}
\end{Snugshade}
\begin{Snugshade}
Q4: \utterance{Under which coaches did Neymar play in Barcelona?}
\end{Snugshade}
%
%
%
\noindent In Q3, no explicit date (e.g., August 2017) is mentioned,
so a challenge is to
detect its temporal nature.
The phrase \phrase{joining PSG} refers to an event (Neymar's transfer to that 
team). 
We could detect this, but have to properly disambiguate it to a normalized date.
The temporal preposition \phrase{before} is a strong cue as well, but 
words like \phrase{before}, \phrase{after}, etc. are also used in non-temporal
contexts; Q2 is an example for this.
Q4 does not seem to be time-dependent at all, when looking at its
surface form. 
However, it is crucial for correct answers that
only coaches are selected whose job periods at FC Barcelona overlap
with that of Neymar. 
Here, detecting the temporal nature is
a big challenge. A second challenge is how to decompose such questions 
 and ensure that the execution contains
an overlap test for the respective time periods.



\noindent \textbf{Approach and Contributions.}
The key idea of this paper is to judiciously decompose
such temporal questions and rewrite the resulting sub-questions
so that they can be separately evaluated by a standard KB-QA system.
The answers for the full questions are then computed by combining and
reasoning on the sub-question results.
For example, Q3 should be decomposed and rewritten into
Q3.1: \utterance{Which teams did Neymar play for?} and
Q3.2: \utterance{When did Neymar join PSG?}.
For the results of Q3.1, we could then retrieve time scopes from the KB,
and compare them with the date returned by Q3.2, using a BEFORE operator.
Analogously, Q4 would require an OVERLAP comparison as a final step.
With the exception of 
the work by~\cite{bao:16}, to which we experimentally
compare our method, we are not aware of any
KB-QA system 
for
such composite questions.

Our solution, called TEQUILA, is built on a rule-based framework that encompasses
four
stages
of processing: (i) detecting temporal questions, (ii) decomposing questions
and
rewriting sub-questions,
(iii) retrieving candidate answers for sub-questions, and
(iv) temporal reasoning to combine and reconcile the results of
the previous stage into final answers.
For stage (iii), we leverage existing KB-QA systems
(state-of-the-art systems QUINT~\cite{abujabal:17} and AQQU~\cite{bast:15}
used in experiments),
that are geared for
answering simple questions.


To the best of our knowledge, this is the first paper
that presents a complete pipeline specific to temporal KB-QA.
Novel contributions also include: (i) a method for decomposing
complex questions, and (ii) the time-constraint-based reasoning
for combining sub-question results into overall answers.
All data and code are public at 
\url{https://github.com/zhenjia2017/tequila},
and a demo is available at \url{https://tequila.mpi-inf.mpg.de/}.


\section{Concepts}
\label{sec:concepts}





In NLP, the markup language TimeML ({\small\tt www.timeml.org}) is widely used
for annotating temporal information in text documents. Our definition
of \emph{temporal questions} is based on two
of its concepts (tags for temporal expressions and temporal signals).

\noindent\textbf{Temporal expressions.} 
\texttt{TIMEX3} tags demarcate four types of temporal expressions. 
Dates and times refer to points in time of different granularities
(e.g., \phrase{May 1, 2010} and  \phrase{9 pm}, respectively).
They occur in fully- or under-specified
forms (e.g., \phrase{May 1, 2010} vs.\ \phrase{last year}).
Durations refer to intervals (e.g., \phrase{two years}), and sets to periodic
events (e.g., \phrase{every Monday}). 
Going beyond TimeML, 
implicit expressions (e.g., \phrase{the Champions League final})
are used
to capture events and their
time scopes~\cite{kuzey:16}. 
Expressions 
can be normalized into standard format (e.g., \phrase{May $2^{nd}$, 2016}
 into \texttt{2016-05-02}). 

\noindent\textbf{Temporal signals.} 
\texttt{SIGNAL} tags mark textual elements that denote 
explicit temporal relations between two TimeML entities (i.e.,
events or temporal
 expressions), such as \phrase{before} or \phrase{during}. 
We extend the TimeML definition to also include cues when an event is
mentioned only implicitly, such as \phrase{joining PSG}.
In addition, we consider {ordinals} like \phrase{first}, \phrase{last}, etc.
These are frequent in questions when entities can be
chronologically ordered, such as \phrase{last} in \utterance{Neymar's last club
before joining PSG}.

%
\noindent{\bf Temporal questions.}
Based on these considerations, we can now \textit{define} a {temporal question} as
any question that contains a temporal expression or
a temporal signal, or whose answer type is temporal.
%
%
%

\noindent{\bf Temporal relations.}
Allen~\cite{allen:90} introduced $13$ temporal relations 
between time intervals for temporal reasoning:
EQUAL, BEFORE, MEETS, OVERLAPS, DURING, STARTS, FINISHES,
and their inverses for all but EQUAL.
However, for an input temporal question,
it is not always straightforward to infer the proper relation.
For example, in Q3 the relation should be BEFORE;  but if we slightly vary
Q3 to:
\begin{Snugshade}
Q5: \utterance{Which team did Neymar play for before joining PSG?},
\end{Snugshade}
\noindent the singular form \phrase{team} suggests that we are
interested in the
MEETS relation, that is, only the last team before the transfer.
%
Frequent trigger words suggesting such relations are, for instance,
 the signals \emph{before},
 \emph{prior to} (for BEFORE or MEETS), \emph{after}, \emph{following} 
 (for {AFTER}), and \emph{during}, \emph{while}, \emph{when}, \emph{in} 
 (for {OVERLAP}).
\section{Method}
\label{sec:tequila}

Given an input question, TEQUILA works in
four stages:
(i) detect if the question is
temporal, 
(ii) decompose the question into simpler 
sub-questions with some form of rewriting, 
(iii) obtain candidate answers and
dates for temporal constraints from a KB-QA system, and
(iv)
apply constraint-based reasoning on the candidates
to produce final answers. 
Our method builds on ideas from the literature on question decomposition
for general QA
~\cite{saquete:09,bao:14,abujabal:17}.
%
Standard NLP tasks 
like POS tagging, NER, and coreference resolution,
are performed on the input question before 
passing it on to TEQUILA.

\subsection{Detecting temporal questions}
\label{subsec:detect}

A question is identified as temporal if it
contains any of the following:
(a) explicit or implicit temporal expressions (dates, 
times, events),
(b) temporal signals (i.e., cue words for temporal relations),
(c) ordinal words (e.g., \textit{first}),
(d) an indication that the answer type is temporal
(e.g., the question starts with \phrase{When}).
We use HeidelTime~\cite{stroetgen:10} 
to tag TIMEX3 expressions in questions.
Named events are identified using a dictionary
curated from Freebase. 
Specifically, if the type of an entity is \phrase{time.event},
its surface forms are
added to the event dictionary.
SIGNAL words and ordinal words are detected using a small dictionary 
as per suggestions from Setzer~\cite{setzer:02}, 
and a list of 
temporal prepositions.
To spot questions whose answers are temporal, 
we use a small set of patterns like
\textit{when}, \textit{what date}, \textit{in what year}, 
and \textit{which century}.

\subsection{Decomposing and rewriting questions}
\label{subsec:decompose}

\begin{table} [t] \small
  \caption{Decomposition and rewriting of questions.
  The \textit{constraint} is the fragment 
  after the SIGNAL word.
  \textit{wh$\ast$} is the question word (e.g., \textit{who}), and 
  $w_i$ are tokens in the question. }
    \begin{tabular}{p{8cm}}
		\toprule
		\textbf{Expected input:} wh$\ast$ $w_1$ $\ldots$ $w_n$ SIGNAL $w_{n+1}$ $\ldots$ $w_p$? \\ \toprule
	    \underline{\textbf{Case 1:} Constraint has both an entity and a relation} 					\\ 
		\textbf{Sub-question 1 pattern:} wh$\ast$ $w_1$ $\ldots$ $w_n$? 						\\
		\textbf{Sub-question 2 pattern:} when $w_{n+1}$ $\ldots$ $w_p$? 						\\
		\textbf{E.g.:} \utterance{where did neymar play before he joined barcelona?} 			\\
		\textbf{Sub-question 1:} \utterance{where did neymar play?} 							\\
		\textbf{Sub-question 2:} \utterance{when neymar joined barcelona?} 						\\ \midrule
		\underline{\textbf{Case 2:} Constraint has no entity but a relation} 									\\ 
		\textbf{Sub-question 1 pattern:} wh$\ast$ $w_1$ $\ldots$ $w_n$? 						\\
		\textbf{Sub-question 2 pattern:} when sq1-entity $w_{n+1}$ $\ldots$ $w_p$? 				\\
		\textbf{E.g.:} \utterance{where did neymar live before playing for clubs?} 				\\
		\textbf{Sub-question 1:} \utterance{where did neymar live?} 							\\
		\textbf{Sub-question 2:} \utterance{when neymar playing for clubs?}						\\ \midrule      
		\underline{\textbf{Case 3:} Constraint has no relation but an entity} 									\\ 
		\textbf{Sub-question 1 pattern:} wh$\ast$ $w_1$ $\ldots$ $w_n$? 						\\
		\textbf{Sub-question 2 pattern:} when $w_{n+1}$ $\ldots$  $w_p$ $w_1$ $\ldots$ $w_n$ ? 	\\
		\textbf{E.g.:} \utterance{who was the brazil team captain before neymar?} 				\\
		\textbf{Sub-question 1:} \utterance{who was the brazil team captain?} 					\\
		\textbf{Sub-question 2:} \utterance{when neymar was the brazil team captain?}			\\ \midrule         
		\underline{\textbf{Case 4:} Constraint is an event name}											\\ 
		\textbf{Sub-question 1 pattern:} wh$\ast$ $w_1$ $\ldots$ $w_n$?							\\
		\textbf{Sub-question 2 pattern:} when did $w_{n+1}$ $\ldots$ $w_p$ happen? 				\\
		\textbf{E.g.:} \utterance{where did neymar play during south africa world cup?} 		\\
		\textbf{Sub-question 1:} \utterance{where did neymar play?} 							\\
		\textbf{Sub-question 2:} \utterance{when did south africa world cup happen?} 			\\ \bottomrule
    \end{tabular}
  \label{tab:decomp-patt}
\end{table}

TEQUILA decomposes a composite temporal question into
one or more {\em non-temporal sub-questions}
(returning candidate answers), and
one or more {\em temporal sub-questions} (returning temporal constraints).
Results of sub-questions are combined by intersecting their
answers. The constraints are applied to time scopes
associated with results of the non-temporal sub-questions.
For brevity, the following explanation focuses on the
case with one non-temporal sub-question, and one
temporal sub-question.
We use a set of 
lexico-syntactic rules (Table~\ref{tab:decomp-patt}) 
designed from
first principles
to decompose and rewrite a question into its components.
Basic intuitions driving these rules are as follows:
\squishlist
	\item The signal word separates the non-temporal and
	temporal sub-questions, acting as a pivot for decomposition;
	\item Each sub-question needs to have an entity and a relation
	(generally represented using verbs) to enable the underlying
	KB-QA systems to handle sub-questions;
	\item If the second sub-question lacks the entity or the relation,
	it is borrowed from the first sub-question;
	\item KB-QA systems are robust to ungrammatical constructs, thus
	precluding the need for linguistically correct sub-questions.
\squishend



%

\subsection{Answering sub-questions}
\label{subsec:answer}

Sub-questions are passed on to the underlying
KB-QA system, which
translates them into SPARQL queries
and executes them on the KB. 
This produces a result set for each
sub-question. Results from the non-temporal sub-question(s) 
are entities of the same type (e.g., football teams). These are  
candidate answers for the full question. With multiple
sub-questions, the candidate sets are intersected.
The temporal sub-questions, on the other hand, return temporal constraints 
such as dates, which act as constraints to filter the non-temporal candidate set.
Candidate answers need to be associated with time scopes,
so that we can evaluate the temporal constraints.
\noindent \textbf{Retrieving time scopes.}
To obtain time scopes, we introduce additional KB lookups; details depend on
the specifics of
the underlying KB.
Freebase, for example, often associates SPO triples with time scopes
by means of
compound value types (CVTs); other KBs may use $n$-tuples ($n>3$)
to attach spatio-temporal attributes to facts.
For example, the Freebase 
predicate
\struct{marriage} is a CVT with attributes including
\struct{marriage.spouse} and \struct{marriage.date}.
When the predicate \struct{marriage.spouse} is used to retrieve answers,
the time scope is retrieved by looking up \struct{marriage.date} in the KB.
On the other hand,
playing for a football club could be
captured in a predicate like 
\struct{team.players}
without temporal information attached, and the job periods are represented as
events in predicates like
\struct{footballPlayer}. \struct{team}. \struct{joinedOnDate} and
\struct{footballPlayer}. \struct{team}. \struct{leftOnDate}).
In such cases, 
TEQUILA considers all kinds of temporal predicates for the
candidate entity, and chooses one based on a {\em similarity measure} between 
the non-temporal predicate (\struct{team.players}) and
potentially relevant temporal predicates 
(\struct{footballPlayer}. \struct{team}. \struct{joinedOnDate},
\struct{footballPlayer.}\struct{award.date}).
%
%
%
The similarity measure is implemented
by selecting tokens in predicate names
(\struct{footballPlayer}, \struct{team}, etc.),
contextualizing the tokens
by
computing \textit{word2vec}
embeddings for them,
averaging
per-token vectors
to get a resultant vector for each predicate~\cite{wieting:16},
and comparing the cosine distance between two predicate vectors. 
The best-matching temporal predicate is chosen for use. When time periods 
are needed
(e.g., for a temporal constraint using OVERLAP),
a pair of begin/end predicates is selected (e.g., 
\struct{footballPlayer}. \struct{team}. \struct{joinedOnDate} and
\struct{leftOnDate}).

\subsection{Reasoning on temporal intervals}
\label{subsec:reason}

%
\begin{table} [t]
  \caption{Temporal reasoning constraints.}
	\resizebox{\columnwidth}{!}{
    \begin{tabular}{l l l }
       \toprule
       \textbf{Relation} & \textbf{Signal word(s)} 	& \textbf{Constraint} 												\\ \toprule
        BEFORE	& \phrase{before}, \phrase{prior to} 		& $end_{ans} \leq begin_{cons}$										\\ \midrule
        AFTER 	& \phrase{after}  					& $begin_{ans} \geq end_{cons}$ 									\\ \midrule
        OVERLAP & \phrase{during}, \phrase{while}, \phrase{when} 		& $begin_{ans} \leq end_{cons} \leq end_{ans}$						\\
		        & \phrase{since}, \phrase{until}, \phrase{in} 		& $begin_{ans} \leq begin_{cons} \leq end_{ans}$					\\
		        & \phrase{at the same time as} 		& $begin_{cons} \leq begin_{ans} \leq end_{ans} \leq end_{cons}$	\\ \bottomrule
    \end{tabular}}
  \label{tab:reason}
\end{table}

For temporal sub-questions, the results are time points,
time intervals,
or sets of dates (e.g., a set of consecutive years during which someone played for
a football team). We cast all these into intervals with start point $begin_{cons}$
and end point $end_{cons}$. These form the temporal constraints against which
we test the time scopes of the non-temporal candidate answers, also cast into
intervals $[begin_{ans}, end_{ans}]$.
The test itself depends on the temporal operator derived from the input question
(e.g., BEFORE, OVERLAP, etc.) (Table~\ref{tab:reason}).
For questions with ordinal constraints (e.g., \textit{last}), 
we sort the (possibly open) intervals to select the appropriate answer.

\section{Experiments}
\label{sec:experiments}

\subsection{Setup}
\label{subsec:setup}

We evaluate TEQUILA on the \textit{TempQuestions} benchmark~\cite{jia:18},
which contains $1,271$ temporal questions 
labeled as questions with explicit, implicit, and ordinal constraints,
and those with temporal answers. 
Questions are paired with their answers over Freebase.
We use three state-of-the-art
KB-QA systems as baselines: AQQU~\cite{bast:15}, QUINT~\cite{abujabal:17}
(code from authors for both), and Bao et al.~\cite{bao:16}
(detailed results from authors). 
The first two are geared for simple questions, 
while
Bao et al. handle complex questions, including temporal ones.
We use TEQUILA as a plug-in for the first two, and directly
evaluate against the system of Bao et al. on $341$ temporal questions from
the ComplexQuestions test set~\cite{bao:16}.
For evaluating baselines, the full question was fed directly to the underlying
system. We report precision, recall, and F1 scores of 
the retrieved answer sets w.r.t. the gold answer sets, and average them
over all test questions.


\subsection{Results and insights}
\label{subsec:results}

\begin{table*} [t] \small
  \caption{Detailed performance of TEQUILA-enabled systems on TempQuestions and ComplexQuestions.}
\newcolumntype{G}{>{\columncolor [gray] {0.90}}c}
    \begin{tabular}{l G G G c c c G G G c c c G G G}
       \toprule     
		\textbf{TempQuestions}				& \multicolumn{3}{G}{\textbf{Aggregate results}} 							& \multicolumn{3}{c}{\textbf{Explicit constraint}}						& \multicolumn{3}{G}{\textbf{Implicit constraint}}						& \multicolumn{3}{c}{\textbf{Temporal answer}}							& \multicolumn{3}{G}{\textbf{Ordinal constraint}}	\\ 
		(1,271 questions)					& \textbf{Prec}			& \textbf{Rec} 				& \textbf{F1}	 		& \textbf{Prec}			& \textbf{Rec} 			& \textbf{F1}			& \textbf{Prec}			& \textbf{Rec}			& \textbf{F1}			& \textbf{Prec}			& \textbf{Rec}			& \textbf{F1}			& \textbf{Prec}			& \textbf{Rec}			& \textbf{F1}			\\ \toprule
		\textbf{AQQU~\cite{bast:15}}		& $24.6$				& $48.0$					& $27.2$		 		& $27.6$ 				& $60.7$				& $31.1$				& $12.9$				& $34.9$				& $14.5$	  			& $26.1$				& $33.5$				& $27.4$				& $28.4$				& $\boldsymbol{57.4}$				& $32.7$				\\
		\textbf{AQQU+TEQUILA}				& $\boldsymbol{36.0}$*	& $42.3$					& $\boldsymbol{36.7}$*		 		& $\boldsymbol{43.8}$*	& $53.8$				& $\boldsymbol{44.6}$*	& $\boldsymbol{29.1}$*	& $34.7$				& $\boldsymbol{29.3}$*	& $27.3$*				& $29.6$				& $\boldsymbol{27.7}$*				& $\boldsymbol{38.0}$*	& $41.3$				& $\boldsymbol{38.6}$*				\\ \midrule
		\textbf{QUINT~\cite{abujabal:17}}	& $27.3$				& $\boldsymbol{52.8}$		& $30.0$		 		& $29.3$ 				& $\boldsymbol{60.9}$	& $32.6$				& $25.6$				& $\boldsymbol{54.4}$	& $27.0$				& $25.2$				& $\boldsymbol{38.2}$	& $27.3$				& $21.3$				& $54.9$	& $26.1$				\\
		\textbf{QUINT+TEQUILA}				& $33.1$*				& $44.6$					& ${34.0}$*		 		& $41.8$* 				& $51.3$				& $42.2$*				& $13.8$				& $43.7$				& $15.7$				& $\boldsymbol{28.6}$*	& $34.5$				& $\boldsymbol{29.4}$*	& $37.0$*				& $42.2$				& ${37.7}$*	\\ \midrule
		\textbf{ComplexQuestions}			& \multicolumn{3}{G}{\textbf{Aggregate results}}							& \multicolumn{3}{c}{\textbf{Explicit constraint}}						& \multicolumn{3}{G}{\textbf{Implicit constraint}}						& \multicolumn{3}{c}{\textbf{Temporal answer}} 							& \multicolumn{3}{G}{\textbf{Ordinal constraint}} 						\\ 
		($341$ questions)						& \textbf{Prec}			& \textbf{Rec} 				& \textbf{F1}	 		& \textbf{Prec}			& \textbf{Rec} 			& \textbf{F1}			& \textbf{Prec}			& \textbf{Rec}			& \textbf{F1}			& \textbf{Prec}			& \textbf{Rec}			& \textbf{F1}			& \textbf{Prec}			& \textbf{Rec}			& \textbf{F1}			\\ \toprule		
		\textbf{Bao et al.~\cite{bao:16}}	& $34.6$				& $48.4$					& $35.9$		 		& $41.1$ 				& $53.2$				& $41.9$				& $26.4$				& $36.5$	 			& $\boldsymbol{27.0}$				& $18.6$				& $\boldsymbol{40.2}$	& $22.3$				& $31.1$				& $\boldsymbol{60.8}$	& $36.1$				\\ \midrule
		\textbf{AQQU~\cite{bast:15}}		& $21.5$				& $50.0$					& $23.3$		 		& $25.0$ 				& $\boldsymbol{60.1}$	& $28.4$				& $11.2$				& $31.2$	 			& $11.4$				& $19.6$				& $35.7$				& $19.2$				& $22.2$				& $54.9$				& $25.3$				\\
		\textbf{AQQU+TEQUILA}				& $\boldsymbol{36.2}$*	& $45.9$					& $\boldsymbol{37.5}$*	& $\boldsymbol{41.2}$* 	& $54.7$				& $\boldsymbol{43.5}$*	& $\boldsymbol{27.5}$*	& $32.6$	 			& $\boldsymbol{27.0}$*	 			& $29.5$*				& $32.1$				& $29.9$*				& $40.2$*				& $45.1$				& $40.8$*				\\ \midrule		
		\textbf{QUINT~\cite{abujabal:17}}	& $22.0$				& $\boldsymbol{50.3}$		& $24.5$		 		& $24.7$ 				& $54.7$				& $27.5$				& $18.8$				& $\boldsymbol{47.9}$	 			& $19.0$				& $16.6$				& $37.5$				& $20.7$				& $20.9$				& $51.3$				& $26.0$				\\
		\textbf{QUINT+TEQUILA}				& $29.6$*				& $44.9$					& $31.1$*		 		& $34.6$* 				& $47.3$				& $36.3$*				& $12.3$				& $42.1$	 			& $13.9$				& $\boldsymbol{33.4}$*	& $37.5$				& $\boldsymbol{33.9}$*	& $\boldsymbol{44.9}$*	& $51.6$*				& $\boldsymbol{45.8}$*	\\ \bottomrule				
    \end{tabular} 
    \\ \raggedright Aggregate results are averaged over the four categories. The highest value in a column for each dataset is in \textbf{bold}. An asterisk (*) indicates statistical significance of TEQUILA-enabled systems over their standalone counterparts, under the $2$-tailed paired $t$-test at $p < 0.05$ level.
  \label{tab:temp-qa-det}
\end{table*}

Results on TempQuestions and the $341$ temporal questions in ComplexQuestions are
shown in Table~\ref{tab:temp-qa-det}. AQQU + TEQUILA and QUINT + TEQUILA
refer to the TEQUILA-enabled versions of the respective baseline systems. We make
the following observations.

\textbf{TEQUILA enables KB-QA systems to answer composite questions with
temporal conditions.}
Overall and category-wise F1-scores show that TEQUILA-enabled systems
significantly outperform the baselines. Note that these systems neither have 
capabilities for handling compositional syntax nor specific support for temporal
questions.
Our decomposition
and rewrite methods are crucial for compositionality, and constraint-based
reasoning on answers is decisive for the temporal dimension.
%
The improvement in F1-scores
stems from a systematic boost in \textit{precision}, 
across most categories. 

\textbf{TEQUILA 
outperforms state-of-the-art baselines.}
Bao et al.~\cite{bao:16} represents the
state-of-the-art in KB-QA, with a generic mechanism for handling constraints in
questions.
TEQUILA-enabled systems outperform Bao et al. on
the temporal slice of ComplexQuestions,
showing that
a tailored method for temporal information needs is worthwhile.
TEQUILA enabled QUINT and AQQU to answer questions like:
\utterance{who is the first husband of julia roberts?},
\utterance{when did francesco sabatini start working on the puerta
de san vicente?}, and
\utterance{who was governor of oregon when shanghai noon was released?}.

\textbf{Error analysis.} Analyzing cases when TEQUILA fails yields insights
towards future work:
%
(i) Decomposition and rewriting were incorrect 
(for example, in \utterance{where did the pilgrims come from before
landing in america?}, \phrase{landing} is incorrectly labeled as a noun, 
triggering case 3 instead of case 1 in Table~\ref{tab:decomp-patt});
(ii) The correct temporal predicate was not found due to limitations of
the similarity function; and
(iii) The temporal
constraint or the time scope to use during reasoning
was wrongly identified. 
%
%

\section{Related Work}
\label{sec:related}

QA has a long tradition in IR and NLP,
including benchmarking tasks in TREC, CLEF,
and SemEval.
This has predominantly focused on retrieving answers
from textual sources. The recent TREC CAR (complex answer retrieval)
resource~\cite{dietz:17}, explores multi-faceted passage answers, but
information needs are still simple. In IBM Watson~\cite{ferruci:12}, 
structured data played
a role, but text was the main source for answers. 
Question decomposition was leveraged, for example, in
~\cite{ferruci:12,saquete:09,yin:15}
for QA over text.
However, re-composition and reasoning over answers works
very differently for textual sources~\cite{saquete:09},
and are not directly applicable for KB-QA.
Compositional semantics of natural language sentences
has been addressed by~\cite{liang:11}
from a general linguistic perspective. Although applicable
to QA, existing systems support only specific cases of composite questions.

KB-QA is a more recent trend,
starting with~\cite{berant:13,cai:13,fader:14,unger:12,yahya:12}.
Most methods have focused on simple questions,
whose SPARQL translations contain only a single
variable (and a few triple patterns for a single set of qualifying entities).
For popular benchmarks like WebQuestions~\cite{berant:13},
the best performing systems use templates and grammars
~\cite{abujabal:17,abujabal:18,bast:15,reddy:14,yin:15},
leverage additional text
~\cite{savenkov:16,xu:16}, or learn end-to-end with extensive training data
~\cite{xu:16,yih:15}. 
These methods do not cope well with complex questions. 
Bao et al.~\cite{bao:16} combined rules with deep learning to address
a variety of complex questions.


\section{Conclusion}
\label{sec:conclusion}
Understanding the compositional semantics of
complex questions is an open challenge in QA.
We focused on temporal question answering over KBs,
as a major step for coping with an important slice of information needs.
Our method showed boosted performance on a recent benchmark, and outperformed
a state-of-the-art baseline on general complex questions.
Our work underlines the value of building reusable modules that
improve several KB-QA systems.

\bibliographystyle{ACM-Reference-Format}
\bibliography{tequila}

\end{document}